\documentstyle[12pt]{article}

\begin{document}
{\sf \begin{center} \noindent
{\Large \bf Non-Riemannian geometrical asymmetrical damping stresses on the Lagrange instability of shear flows}\\[3mm]

by \\[0.3cm]

{\sl L.C. Garcia de Andrade}\\

\vspace{0.5cm} Departamento de F\'{\i}sica
Te\'orica -- IF -- Universidade do Estado do Rio de Janeiro-UERJ\\[-3mm]
Rua S\~ao Francisco Xavier, 524\\[-3mm]
Cep 20550-003, Maracan\~a, Rio de Janeiro, RJ, Brasil\\[-3mm]
Electronic mail address: garcia@dft.if.uerj.br\\[-3mm]
\vspace{2cm} {\bf Abstract}
\end{center}
\paragraph*{}
It is shown that the physical interpretation of Elie Cartan
three-dimensional space torsion as couple asymmetric stress, has the
effect of damping, previously Riemannian unstable Couette planar
shear flow, leading to stability of the flow in the Lagrangean
sense. Actually, since the flow speed is inversely proportional to
torsion, it has the effect of causing a damping in the planar flow
atenuating the instability effect. In this sense we may say that
Cartan torsion induces shear viscous asymmetric stresses in the
fluid, which are able to damp the instability of the flow. The
stability of the flow is computed from the sectional curvature in
non-Riemannian three-dimensional manifold. Marginal stability is
asssumed by making the sectional non-Riemannian curvature zero,
which allows us to determine the speeds of flows able to induce this
stability. The ideas discussed here show that torsion plays the
geometrical role of magnetic field in hydromagnetic instability of
Couette flows recently investigated by Bonnano and Urpin (PRE,
(2007,in press) can be extended and applied to plastic flows with
microstructure defects. Recently Riemannian asymmetric stresses in
magnetohydrodynamics (MHD) have been considered by Billig
(2004).\vspace{0.5cm} \noindent {\bf PACS numbers:}
\hfill\parbox[t]{13.5cm}{02.40.Hw:Riemannian geometries}

\newpage
\newpage
 \section{Introduction}
 Recently two papers on the use of stresses magnetic\cite{1,2} or nonmagnetic \cite{3} have been considered by
 Billig \cite{1}, Bonnano and Urpin \cite{2} and L'ov et al \cite{3}. In the first the Riemannian geometry was used along asymmetric
 stresses in MHD equations. In the second, Bonnano et al showed that the presence of magnetic field could induce
 instabilities in Couette plane flows , and in the third also instabilities in polymer interactions of fluids were investigated
 also in MHD. All these three papers together motivated us to use Elie Cartan idea \cite{1} of associating a
 non-Riemannian asymmetric  connection known nowadays by the name of Cartan torsion tensor \cite{5,6,7}, to moment or torque stresses to a Couette plane
 flow which has been systemmatically used over the
 years to investigate its Lagrangean stability. Cartan torsion has been also used in the fields of gravitation and cosmology, where
 torsion is in general associated with a spin density, often called Einstein-Cartan (EC) theory \cite{8}, or in modern language of
 theoretical physics, a torsioned high-energy
 menbranes \cite{9}. Analogies between torsion in defects and in solids \cite{10} and EC gravity has been put foward by
 Maugin \cite{11} and Kr\"{o}ner \cite{12} and developed most recently by Epstein \cite{13}. Despite Cartan
 first motivation to apply torsion to physics was Einstein general relativistic cosmology, he argued that this study could be done
 easily by investigating the mechanical torque stresses \cite{4} in three-dimensional, which
could be exactly due to asymmetries in coupled shear stresses. In
this brief report, we give an example that
 torsion can be applied also in physics of fluids, when the fluid, in the case a Couette type flow, is viscous and sheared.
 Actually Kambe \cite{14} has previously investigate Couette planar flows stability, and showed that Riemann curvature tensor
 would exist in the case of existence of pressure, even a constant one. The stability of the Couette planar flow is also
 obtained by him, using the symmetry in the case of symmetric scalar stress potential and by making use of the technique
 of Ricci sectional curvature \cite{15}, where the negative sectional curvature indicates instability of the flow, in the
 Lagrangean sense. This idea can be easily shifted to flows in the case they have strong shear and viscosity and are able to support
 these antisymmetric stresses. In this brief report, we also show that allowing the presence of shearing stresses, and associating
 torsion to these stresses, the Couette flows are can be stable even
 in the Lagrangean sense. This can be explained by the fact, that as
 we shall show, velocities involved in the fluid are inversely
 proportional to the flow torsion, and as torsion grows the velocity
 in the fluid decreases, which is a damping like physical effect in
 the flow. Actually, Kondo \cite{10} has shown that the antisymmetry of stress tensor derivativative is associated with Cartan
 torsion. To test the range of this Lagrangean stability, we assume a marginal stability and put the sectional
 curvature for the torsioned connection to zero, and find an algebraic equation to previously arbitrary velocities in the
 fluid. Torsion would only allows for the instability if its sectional curvature is negative. A previously application of
 a non-torsion free connection in fluids, in the context of solitons, have been investigated by Ricca \cite{16}.
 Rakotomanana \cite{17} in turn have investigated NR manifolds with Cartan torsion, in the context of the geometrical approach to the thermodynamics of
 dissipating continua. The non-Riemannian geometry of vortex acoustic flows have been also recently addressed \cite{18}.
 The paper is organized as follows: Section II presents a brief review of non-Riemannian geometry in the coordinate free
 language. Section III presents model, along with computation of the
 Lagrange stability of non-Riemannian Couette
 flow. Section IV presents the conclusions.
 \section{Sectional non-Riemannian curvature}
 In this section, before we add we make a brief review of the differential geometry of surfaces in coordinate-free language.
 The Riemann curvature is defined by
 \begin{equation}
 R(X,Y)Z:={\nabla}_{X}{\nabla}_{Y}Z-{\nabla}_{Y}{\nabla}_{X}Z-{\nabla}_{[X,Y]}Z\label{1}
 \end{equation}
where $X {\epsilon} T\cal{M}$ is the vector representation which is
defined on the tangent space $T\cal{M}$ to the manifold $\cal{M}$.
Here ${\nabla}_{X}Y$ represents the covariant derivative given by
\begin{equation}
{\nabla}_{X}{Y}= (X.{\nabla})Y\label{2}
 \end{equation}
which for the physicists is intuitive, since we are saying that we
are performing derivative along the X direction. The expression
$[X,Y]$ represents the commutator, which on a vector basis frame
${\vec{e}}_{l}$ in this tangent sub-manifold defined by
\begin{equation}
X= X_{k}{\vec{e}}_{k}\label{3}
\end{equation}
or in the dual basis ${{\partial}_{k}}$
\begin{equation}
X= X^{k}{\partial}_{k}\label{4}
\end{equation}
can be expressed as
\begin{equation}
[X,Y]= (X,Y)^{k}{\partial}_{k}\label{5}
\end{equation}
In this same coordinate basis now we are able to write the curvature
expression (\ref{1}) as
\begin{equation}
R(X,Y)Z:=[{R^{l}}_{jkp}Z^{j}X^{k}Y^{p}]{\partial}_{l}\label{6}
\end{equation}
where the Einstein summation convention of tensor calculus is used.
The expression $R(X,Y)Y$ which we shall compute bellow is called
Ricci curvature. The sectional curvature which is very useful in
future computations is defined by
\begin{equation}
K^{Riem}(X,Y):=\frac{<R(X,Y)Y,X>}{S(X,Y)}\label{7}
\end{equation}
where $S(X,Y)$ is defined by
\begin{equation}
{S(X,Y)}:= ||X||^{2}||Y||^{2}-<X,Y>^{2}\label{8}
\end{equation}
where the symbol $<,>$ implies internal product. In the
non-Riemannian (NR) case, the torsion two-form $T(X,Y)$ is defined
by
\begin{equation}
T(X,Y):=
\frac{1}{2}[\bar{{\nabla}}_{X}Y-\bar{{\nabla}}_{Y}X-[X,Y]]\label{9}
\end{equation}
where $\bar{\nabla}$ is the non-Riemannian connection \cite{7}
endowed with torsion. As in EC theory \cite{5} the geodesic equation
does not depend on torsion; only Jacobi deviation equation depends
on torsion which is enough for investigate the role of torsion on
stability. Since the Jacobi equation is given by
\begin{equation}
\frac{d^{2}J}{ds^{2}}=[||{\nabla}_{\vec{t}}\vec{e}_{J}||^{2}-K^{NR}(t,\vec{e}_{J})||J||\label{10}
\end{equation}
where $||\vec{e}_{J}||=1$ and J is the Jacobi field, representing
the separation between geodesics,while $\vec{t}$ is the geodesic
tangent vector. Here $K^{NR}(X,Y)$ is given by
\begin{equation}
K^{NR}(X,Y)= K^{Riem}(X,Y)+2<T(X,Y),\bar{\nabla}_{Y}X>\label{11}
\end{equation}
Here as we shall see bellow the geodesic equation is
${\nabla}_{Y}Y=0$ simplified this expression. Note from this
expression that the instability, or separation of the geodesics in
the flow
\begin{equation}
\frac{d^{2}J}{ds^{2}}\ge{0}\label{12}
\end{equation}
implies that $K^{Riem}<0$ which is the condition for Lagrange
instabillity.
\section{Couette shear flow stability in non-Riemannian background} In this
section we shall consider the Couette constant pressure, planar
shear flow \cite{11}
\begin{equation}
Y=(U(y),0,0) \label{13}
\end{equation}
with constant pressure p where $U(y)=y$ and X is given by
\begin{equation}
X=\vec{v}_{l}e_{l}\label{14}
\end{equation}
where $e_{l}:= exp[i(\vec{l}\vec{x})]$ and $\vec{x}=(x,y,z)$ and
$\vec{l}:=(l_{x},l_{y},l_{z})$ is the wave number. Here
$\vec{v}_{l}$ represents an arbitrary velocity, which shall be
determined below in order to generate non-Riemannian stability of
Couette shear flows. The hydrodynamics in ${\cal{R}}^{3}$ Euclidean
space \cite{12} is given by
\begin{equation}
\bar{\nabla}_{X}Y=(X.{\nabla})Y+grad{p_{XY}} \label{15}
\end{equation}
where the covariant derivative on the RHS of this equation \cite{12}
is
\begin{equation}
(X.{\nabla})Y=e_{l}(\vec{v_{l}}.{\nabla})(U(y),0,0)=e_{l}({v^{y}}_{l}U'(y),0,0)
\label{16}
\end{equation}
A  simple computation led Kambe to the result
\begin{equation}
{\nabla}^{2}p_{XY}={\nabla}^{2}p_{YX}=-il_{x}{v_{l}}^{y}e_{l}\label{17}
\end{equation}
since $U'(y)=1$. Note however that the equation (\ref{17}) does not
necessarily implies that $p_{XY}=p_{YX}$, since this is a sufficient
but not necessary solution in the mathematical language. This can be
easily seen by the argument that the equation (\ref{15}) is
equivalent to
\begin{equation}
{\nabla}^{2}[p_{XY}-p_{YX}]=0
 \label{18}
\end{equation}
and since ${\nabla}^{2}={\nabla}.{\nabla}$ we have that
\begin{equation}
{\nabla}p_{XY}-{\nabla}p_{YX}=\vec{c} \label{19}
\end{equation}
where $\vec{c}$ is an arbitrary constant. As we shall show this
reasoning leads exactly to Cartan torsion 2-form $T(X,Y)$ which
using the expression
\begin{equation}
p_{XY}=-il_{x}{v_{l}}^{y} \label{20}
\end{equation}
implies that Cartan torsion vector can be expressed as
\begin{equation}
T(X,Y)=grad(p_{XY}-p_{YX})=
[{T^{k}}_{21}X^{2}Y^{1}]{\partial}_{k}\label{21}
\end{equation}
where the covariant components of torsion are
\begin{equation}
{T^{k}}_{21}=\frac{c}{{v_{x}}^{l}U(y)}=
\frac{c}{{v_{x}}^{l}y}\label{22}
\end{equation}
An example of totally skew-torsion \cite{19} has shown also be
presented in the cholesteric blue phase of liquid crystals. The
torsion vector (\ref{22}) in the NR sectional curvature obtains
\begin{equation}
2<T(X,Y),\bar{\nabla}_{Y}X>=
{(2{\pi})^{3}}{U'}^{2}\frac{{m^{2}}_{x}}{m^{2}}|{v_{y}}^{m}|^{2}
\label{23}
\end{equation}
which comes from the relation
\begin{equation}
2<T(X,Y),\bar{\nabla}_{Y}X>=K^{Riem}(X,Y)
{(2{\pi})^{3}}{U'}^{2}\frac{{m^{2}}_{x}}{m^{2}}|{v_{y}}^{m}|^{2}
\label{24}
\end{equation}
which is the condition for marginal stability of
\begin{equation}
K^{NR}(X,Y)=0 \label{25}
\end{equation}
Making use of the expression for the covariant derivative in the
Riemann-Cartan connection $\bar{\nabla}$
\begin{equation}
\bar{\nabla}_{Y}X=
im_{x}e_{m}y\vec{v_{m}}-\frac{m_{x}}{m^{2}}e_{m}\vec{m}\label{26}
\end{equation}
Substitution of the value of torsion form above and this covariant
derivative into expression (\ref{23}) yields
\begin{equation}
im_{x}U(y)e_{m}U(y)<\vec{v_{m}},\vec{c}>-\frac{m_{x}}{m^{2}}{v_{m}}^{y}U'(y)e_{m}<\vec{m},\vec{c}>=
{(2{\pi})^{3}}{U'}^{2}\frac{{m^{2}}_{x}}{m^{2}}|{v_{y}}^{m}|^{2}
\label{27}
\end{equation}
Thus since $U(y)=y$ this equation is simply solved when we choose a
direction for $\vec{v_{m}}$. The simplest choice is
$\vec{c}=c\vec{m}$, where $<\vec{m},\vec{m}>=m^{2}$. Since \cite{12}
$<{v}_{m},\vec{m}>=0$, which upon substitution into (\ref{27})
yields
\begin{equation}
{v_{m}}^{y}=-\frac{{m^{2}}c}{(2{\pi})^{3}} \label{28}
\end{equation}
Physically this means that the curvature acts as a damping to the
flow. Substitution of the value of the torsion scalar c into the
value of torsion component one obtains
\begin{equation}
{T^{k}}_{21}=-\frac{1}{2}\frac{(2{\pi})^{3}{v^{y}}_{m}}{m^{2}{v^{l}}_{x}y}
\label{29}
\end{equation}
A simple observation of this expression shows that the torsion has
the proper units of $\sim{{L}^{-1}}$ where L represents the length
scale in the flow or liquid crystal.
\section{Conclusions}
One of the most important features of the investigation of the
stability of flows in the Euclidean manifold ${\cal{E}}^{3}$, in
some detail the stability of an incompressible or volume preserving
flow, using the method of the sign of the Ricci sectional curvature.
Also important is the issue of stability in plasma astrophysics as
well as in fluid mechanics. In this paper we discuss and present the
contribution of Cartan torsion tensor on damping the Riemannian
Lagrange instabilities of viscous Couette shear flows. We could say
that contrary to the Riemannian case \cite{12} where the flow is
unstable in the Lagrangean particle sense, but is neutrally stable,
here the Couette shear flow may be fully stable for an appropriate
chice of velocities. The role of torsion in crystal curvature
frustration \cite{20} could also be investigate concerning the
stability of geodesics.


\newpage
\begin{thebibliography}{18}
\bibitem{1} Y.Billig, Magnetic hydrodynamics with asymmetric stress tensor, ArXiv:math-ph/0401052v1.
\bibitem{2} A. Bonnano, Y. Urpin, Phys Rev E (2007) in press.
\bibitem{3} V. S. L'vov, A. Polyalov, I. Procaccia,V. Tiberkevich,
Phys. Rev. E 71 (2005) 016305.
\bibitem{4} E. Cartan, C.R. Acad. Sciences 174 (Paris) (1922) 593.
L.C. Garcia de Andrade, C. A. Souza Lima jr, On the differential
geometry of torque stresses in cylindrical solids (1995) Extract
Mathematicae 11, 1, 59.
\bibitem{5} E. Cartan, A. Einstein, Letters on Absolute
Parallelism-(1929-1932), ed. R. Debever,Princeton (1979).
\bibitem{6} E. Cartan, Riemann Spaces (2000) MIT Press, Boston.
\bibitem{7} E. Cartan, Riemannian Geometry on Orthonormal Frame
(2000) Princeton.
\bibitem{8} F. W. Hehl, Yu N. Obukhov, Foundations of Classical
Electrodynamics: Charge, Flux and Metric (2003), Birkhauser.
\bibitem{9} B. Muhopadhyaya, S. Sen, S, SenGupta, Phys Rev Lett. 89
(2002) 121101.
\bibitem{10} K. Kondo, RAAG Memoirs of the Unified Problems in the
science and Engineering by Means of Geometry (1955) vol. I.
\bibitem{11} G. Maugin, Material Inhomogeneous in Elasticity.
\bibitem{12} E. Kroener, Theory of Defects, Les Houches school,
(1970) Dunod, Paris.
\bibitem{13} M. Epstein, M. Elzanowski, Material Inhomogeneities
and their Evolution: A Geometric Approach (2007) Springer in press.
\bibitem{14} T. Kambe, Geometrical Theory of Dynamical Systems and
Fluid Flows (2004) World Scientific.
\bibitem{15} I. Benn, W. Tucker, An introduction to spinors and
geometry with applications to physics (1987) Adam Hilger, Bristol.
\bibitem{16} R. Ricca, Phys Rev A 43 (1990) 4281.
\bibitem{17} L.R. Rakotomanana, A Geometric Approach to the Thermomechanics of
continua (2003) Birkhauser.
\bibitem{18} L.C. Garcia de Andrade,Phys Rev D 70 (2004) 640004-1
\bibitem{19} E. Dubois-Violette, E. Pansu, Which Universe is Blue Phase, in Geometry in Condensed Matter Physics (1990)
World Scientific.
\bibitem{20} J.F. Sadoc, R. Mosseri, Geometric Frustration (2006) Cambridge university press.
,\end{thebibliography}
\end{document}